\documentclass[12pt]{article}
\pdfoutput=1

\usepackage{amsmath}
\usepackage{amssymb}
\usepackage{graphicx}
\usepackage{cite}
\usepackage{multirow}
\usepackage{longtable}
\usepackage{lscape}

\allowdisplaybreaks[4]  % for page break

\newcommand{\beq}{\begin{equation}}
\newcommand{\eeq}{\end{equation}}
\newcommand{\be}{\begin{eqnarray}}
\newcommand{\ee}{\end{eqnarray}}

\newcommand{\ts}{\sin^2 2\theta}

\newcommand{\Sec}{Sec.}
\newcommand{\capdef}{}
\newcommand{\mycaption}[2][\capdef]{\renewcommand{\capdef}{#2}%
       \caption[#1]{{\footnotesize #2}}}
\makeatletter
\renewcommand{\fnum@table}{\textbf{\tablename~\thetable}}
\renewcommand{\fnum@figure}{\textbf{\figurename~\thefigure}}
\makeatother
\begin{document}

\begin{titlepage}

\renewcommand{\thefootnote}{\alph{footnote}}

\vspace*{-3.cm}
\begin{flushright}
% Report numbers
EURONU-WP6-11-32  \\
IFIC/11-22
\end{flushright}

\vspace*{0.2cm}

\renewcommand{\thefootnote}{\fnsymbol{footnote}}
\setcounter{footnote}{-1}

{\begin{center} {\large\bf
 Short-baseline Neutrino Oscillation Waves in Ultra-large Liquid Scintillator Detectors   \\[0.3cm]
    } \end{center}} 

\renewcommand{\thefootnote}{\alph{footnote}}

\vspace*{.3cm}
% \vspace*{.3cm}
{\begin{center}
{\large \today}
\end{center}
}

{\begin{center} {
                \large{\sc
                 Sanjib~Kumar~Agarwalla\footnote[1]{\makebox[1.cm]{Email:}
                 Sanjib.Agarwalla@ific.uv.es}},
                 \large{\sc
                 J.M. Conrad\footnote[2]{\makebox[1.cm]{Email:}
                 conrad@mit.edu}},
                 \large{\sc
                 M.H. Shaevitz\footnote[3]{\makebox[1.cm]{Email:}
                 shaevitz@nevis.columbia.edu}}
                 }
\end{center}
}
\vspace*{0cm}
{\it
\begin{center}

\footnotemark[1]
       Instituto de F\'{\i}sica Corpuscular, CSIC-Universitat de Val\`encia, \\
       Apartado de Correos 22085, E-46071 Valencia, Spain \\
\footnotemark[2]
       Department of Physics, Massachusetts Institute of Technology, \\
       Cambridge, Massachusetts 02139, USA \\
\footnotemark[3]
       Department of Physics, Columbia University, New York, \\
       New York 10027, USA

\end{center}}

\vspace*{0.5cm}

{\Large \bf
\begin{center} Abstract \end{center}  }

Powerful new multi-kiloton liquid scintillator neutrino detectors,
including NO$\nu$A and LENA, will come on-line within the next decade.
When these are coupled with a modest-power decay-at-rest (DAR)
neutrino source at short-baseline, these detectors can decisively
address the recent ambiguous signals for neutrino oscillations at high $\Delta m^2$.  
These detectors are $>$~50 m long, and so with a DAR beam, the characteristic oscillation wave will be
apparent over the length of the detector, providing a powerful
verification of the oscillation phenomena. LENA can simultaneously
perform $\bar \nu_\mu \to \bar \nu_e$ appearance and
$\nu_e \to \nu_e$ disappearance searches with unprecedented sensitivity.
NO$\nu$A is likely limited to $\nu_e$ disappearance given its present design,
but also has excellent sensitivity in the high $\Delta m^2$ region.
For the appearance channel, LENA could provide a stringent test of the LSND and 
MiniBooNE signal regions at $> 5\,\sigma$ with a reduced fiducial volume of 5 kt and a 10 kW
neutrino source. In addition, the LENA and  NO$\nu$A disappearance sensitivities in $\nu_e$ 
mode are complementary to the recent reactor anomaly indicating possible $\bar \nu_e$ 
disappearance and would cover this possible oscillation signal at the $3\,\sigma$ level.

\vspace*{.5cm}

\end{titlepage}

%----------------------------------------------------------
\newpage

\renewcommand{\thefootnote}{\arabic{footnote}}
\setcounter{footnote}{0}

\section{Introduction}
\label{sec:intro}

Recent results from short-baseline (SBL) neutrino oscillation studies seem to point towards 
the existence of sterile neutrinos.  The strongest indication comes from the LSND
experiment~\cite{Athanassopoulos:1995iw,Athanassopoulos:1996jb,Athanassopoulos:1997er,Aguilar:2001ty},
which has reported a $3.8\,\sigma$ excess of $\bar\nu_e$ events in a
beam of $\bar\nu_\mu$.  The associated $\Delta m^2$ of the assumed
oscillation is too large to be explained with only three active
neutrinos, and so oscillations involving sterile neutrinos are
invoked which do not couple to W and Z bosons. The LSND result is supported by an apparent 
excess of $\bar\nu_e$ events in a beam of $\bar\nu_\mu$ above
$475\,\mathrm{MeV}$ observed in MiniBooNE
~\cite{AguilarArevalo:2010wv}.  This result is consistent with
two-neutrino $\bar\nu_{\mu} \rightarrow \bar\nu_e$ oscillations at
99.4\% confidence level. 

More motivation has arisen from a recent reanalysis of the expected
$\bar \nu_e$ flux emitted from nuclear reactors ~\cite{Mueller:2011nm}
that leads to an observed deficit of $\bar{\nu}_e$ at 98.6\% C.L.  The
reactor anti-neutrino flux prediction depends on accurate prediction of
the reactor isotopes produced as a function of time convoluted with
the spectra of anti-neutrinos produced by each isotope
\cite{Bemporad:2001qy}.  This anomaly arises from new calculations of
the second ingredient, the anti-neutrino spectra, that update analyses
from the 1980's \cite{Schreckenbach:1985ep, VonFeilitzsch:1982jw,
  Hahn:1989zr} by including the latest information from nuclear
databases and improving the calculational techniques
\cite{Mueller:2011nm}.  The overall reduction in predicted flux
compared to the existing data from SBL neutrino
experiments can be interpreted as oscillations at baselines of order
10--100~m~\cite{Mention:2011rk} consistent with the LSND and MiniBooNE
anti-neutrino results which require a mass squared difference of the
order 0.1--10 eV$^2$.

The SBL experimental results can be described using models which involve
three active and one sterile (3+1) or two sterile (3+2) neutrinos
\cite{Peres:2000ic,Strumia:2002fw,Grimus:2001mn,Sorel:2003hf,Maltoni:2002xd,
Karagiorgi:2006jf,Maltoni:2007zf,Acero:2007su,Giunti:2009zz,Giunti:2010jt},
with relatively small mixing to the active flavors, allowing for oscillations with 
high $\Delta m^2$. While a sterile neutrino is theoretically well motivated
with a large number of phenomenological consequences, the actual oscillation parameters
required by the LSND and MiniBooNE anti-neutrino data are in conflict with various
constraints imposed by other SBL neutrino oscillation experiments, most notably 
CDHS~\cite{Dydak:1983zq}, MiniBooNE disappearance measurements~\cite{AguilarArevalo:2009yj} 
and also the data from reactor experiments, 
like Bugey~\cite{Declais:1994su}. Also, there is no sign of sterile oscillations in
atmospheric and solar neutrino data in the required parameter range~\cite{Maltoni:2007zf}.     
This tension could be relieved by fitting neutrino and anti-neutrino
results separately~\cite{Karagiorgi:2009nb}, suggesting that neutrinos
might oscillate differently than anti-neutrinos. When analyzed within
the context of the recent reactor anti-neutrino anomaly, a low, but
acceptable, compatibility is found in the global fit using a (3+2)
model~\cite{Kopp:2011qd}.   The oscillation parameters found in these 
fits have $\Delta m^2 \sim 1$ eV$^2$ associated with the 
sterile neutrino oscillations

In the next five years, a series of small experiments will explore the
(3+1) and (3+2) signal space further.  This program includes continued
MiniBooNE running in anti-neutrino mode and running of the MicroBooNE
Experiment in neutrino mode \cite{Chen:2007zz}.  Also, the upcoming
KATRIN beta decay experiment is sensitive to a light sterile
neutrino~\cite{Barrett:2011jg}. The Low Energy Neutrino Spectroscopy
(LENS) detector, which is now in the prototype stage, may also run
with radioactive neutrino sources in order to explore the question of
sterile neutrinos \cite{Grieb:2006mp,Agarwalla:2010gd}.  However, none
of these experiments are expected to provide confirmation at the
5~$\sigma$ confidence level in the near future.

The next major step in the search for SBL oscillations can be achieved
by pairing one of the ultra-large liquid scintillator detectors
planned for the near future with a low energy neutrino beam.  These
experiments can study $\bar \nu_\mu \rightarrow \bar \nu_e$ and $\nu_e
\rightarrow \nu_e$ oscillations. The high-sensitivity $\nu_e$
disappearance search will bring unique information to the fits for
comparison to the reactor $\bar \nu_e$ disappearance results.  A
relatively low power neutrino source producing neutrinos via the pion
decay-at-rest (DAR) chain, is ideal.  The DAR chain leads to a beam dominated by
neutrinos between 20 and 52 MeV, with a well-defined flavor content of
$\nu_e$, $\nu_\mu$ and $\bar \nu_\mu$, as shown in
Fig.~\ref{fig:flux}. This source may be provided by a low energy
proton accelerator with a beam impinging on a target/dump.
Potentially, this can be a prototype for the cyclotrons planned for
the DAE$\delta$ALUS $CP$-violation 
search~\cite{Conrad:2009mh,Agarwalla:2010nn,Alonso:2010fs}.  
This would be a small, relatively inexpensive
proton source which can be easily positioned within 20 m of the
detectors.

The energy range of the DAR beam is well suited to observe the
$L/E$ dependence of the oscillation wave across the length scales of
presently planned detectors.  A precise search for appearance and disappearance oscillations 
can be achieved by fitting the observed events with respect to expectation as a function of $L/E$.  
Besides observing oscillations from a simple counting analysis, these types of experiments also will confirm 
that oscillations are taking place by seeing the $L/E$ variations within the detector.  
For a 40 MeV neutrino energy and $\Delta m^2$ = 2 eV$^2$, the characteristic oscillation length 
is $L_{\rm osc} \simeq$ 50 m which matches with the length of the liquid scintillator detectors 
being proposed for the near-future program.  Because of the 1/$L^2$ falloff of the neutrino flux with distance, 
the part of the detector within 50 m of the DAR source provides the main oscillation search sensitivity.  
Therefore, maintaining at least 50 m of detector length near the source appears to be best for covering 
the oscillation space of interest.

In this paper, we consider two examples of scintillator detectors.
Our example of an unsegmented detector is LENA~\cite{Wurm:2011zn},
under consideration within the LAGUNA~\cite{Angus:2010sz} project and the
forthcoming LAGUNA-LBNO~\cite{lbno} design studies. 
The conclusions should be similar for other unsegmented detectors 
such as Hanohano\cite{Learned:2007zz,KYLearned:2009rv}.
Our example of a finely segmented detector is
NO$\nu$A~\cite{Ayres:2007tu}, which is under construction for
long-baseline studies using a beam from Fermilab.  This detector is
found to be less powerful than LENA, but has the advantage that it is
already under construction and has an ideal space for the neutrino
source.  Each of these experiments can be used as designed, with only
the addition of the beam source.

We note that the large liquid scintillator detectors represent one of
three types of future neutrino detectors under discussion.  The other
two types of detectors are water and liquid argon (LAr) based detectors.
Ref.~\cite{Agarwalla:2010zu} provides an initial exploration of SBL oscillation 
searches using a DAR neutrino beam and the existing Super-Kamiokande detector
with the addition of Gadolinium. The opportunities for an LAr detector will 
be explored in~\cite{future}.

This paper is organized as follows. We begin with the description of neutrino source and
DAR flux in \Sec~2. In \Sec~3, we deal with the relevant SBL neutrino oscillation probabilities.
After that, we describe the characteristics of liquid scintillator detectors (NO$\nu$A and LENA)
in detail in \Sec~4. We also discuss the possible charged current (CC) interactions of a DAR beam 
in liquid scintillator. In \Sec~5, we present our results for sterile oscillation searches using 
DAR-LENA and DAR-NO$\nu$A setups. We summarize and draw our conclusions in \Sec~6.  

%%%%%%%%%%%%%%%%%%%%%%
\section{The Neutrino Source and DAR Flux}
%%%%%%%%%%%%%%%%%%%%%%

%%%%%%%%%%%%%%%%%%%%%%%%%%%%%%%%%%%%%%%%%%%%%%%%%%%%%%%%
\begin{figure}[tp]
\centering
\includegraphics[width=0.5\textwidth]{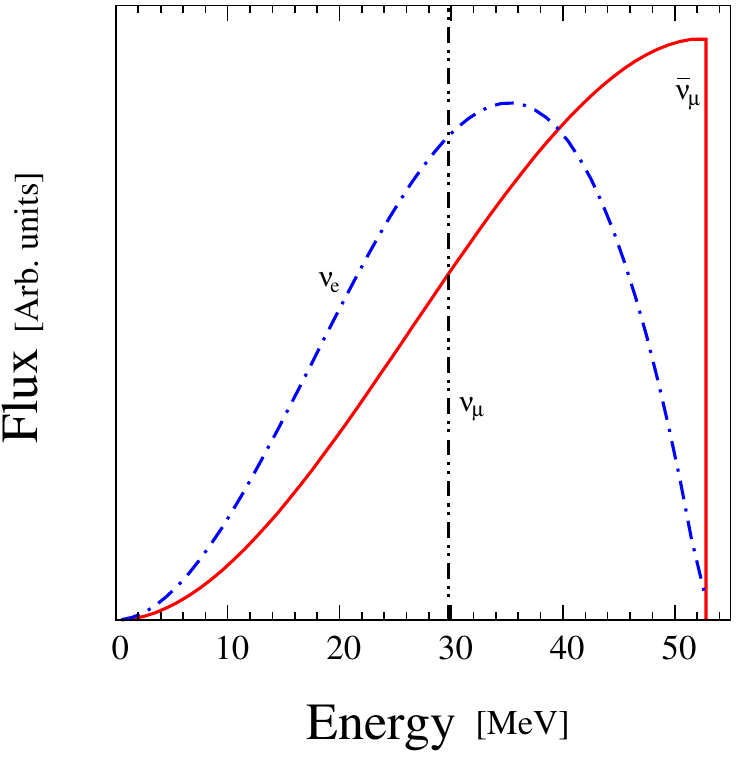}
\mycaption{\label{fig:flux} The energy distribution of different flavor of neutrinos in a DAR beam.}
\end{figure}
%%%%%%%%%%%%%%%%%%%%%%%%%%%%%%%%%%%%%%%%%%%%%%%%%%%%%%%%%

%%%%%%%%%%%%%%%%%%
\begin{table}[t]
\begin{center} 
  \begin{tabular}{ccc|ccc|ccc}
    \hline
    $4\times 10^{21}$ per year, per flavor ($\nu_\mu$, $\bar \nu_\mu$ and $\nu_e$), \\ 
    $1.6 \times 10^{18}$ per year of $\bar \nu_e$ ($4\times 10^{-4}$ compared to other flavors);\\
    Delivered as 100 kW average power, with 200 kW instantaneous power, \\ 
    (50\% duty factor allowing equal beam-on and beam-off data sets); \\
    800 MeV protons on target;\\
    $\pm$25 cm smearing (assumed flat) on neutrino production point;\\
    20 m distance from average production point to face of detector fiducial region.\\
    \hline
  \end{tabular}
  \caption{Characteristics of the neutrino source assumed in calculations,
unless explicitly noted in the text.
  \label{tab:beamparams}}
\end{center}
\end{table}
%%%%%%%%%%%%%%%%%%

In a stopped pion source a proton beam of $\sim 1$ GeV energy
interacts in a low-A target producing $\pi^+$ and, at a low level,
$\pi^-$ mesons. The pions then are brought to rest in a high-A beam
stop.  The $\pi^-$ will capture. The $\pi^+$ will produce the
following cascade of decays
%%%%%%%%%%%
\begin{eqnarray*}
\pi^+ &\rightarrow &\mu^+ +\nu_\mu \\
&& \hspace{0.1cm}\raisebox{0.5em}{$\mid$}\!\negthickspace\rightarrow\ e^+ + \nu_e +\bar\nu_\mu 
\end{eqnarray*}
%%%%%%%%
resulting in $\nu_\mu$, $\bar\nu_\mu$ and $\nu_e$, but no $\bar\nu_e$,
as shown in Fig.~\ref{fig:flux}.  The resulting flux is isotropic.

Certain choices of proton beam kinetic energy and neutrino beam stop
design are crucial to producing the purest, highest rate DAR flux
\cite{Allen:1989dt, Burman:1989dq}.  As
discussed in Ref.~\cite{Alonso:2010fs}, the production of DAR
neutrinos is relatively flat and maximized in the region of beam
kinetic energy 700 to 1500 MeV for a given beam power. Selecting the
lower energy range reduces the $\pi^-$ production, which has a
threshold of 602 MeV.  Above 1500 MeV, kaon production is above
threshold.  It is for this reason that high energy proton beams, such
as the 3 GeV beam proposed for Project X~\cite{Tschirhart:2011zza}, are
poor sources for a DAR beam.  The most efficient DAR production is on
a light target which has tightly bound neutrons.  In high-A targets,
energy is wasted in neutron production~\cite{spallation}.  High-A
targets also have increased $\pi^-$ production compared to light
targets \cite{Burman:1996gt}. For this reason, a spallation neutron
source, which uses high-A targets for neutron production, is not an
ideal venue for the kind of experiments described here.  In the optimized DAR
beam designs, the light target for production is typically embedded in
a high-A, dense material. This increases the probability that the
$\pi^-$ produced in the light target will be captured before
decay-in-flight (DIF), minimizing the decay chain that produces $\bar
\nu_e$ backgrounds.  Upstream targets, as are used in neutron
production and also for isotope production, must not be used for this
study, because these will substantially increase DIF backgrounds
\cite{Athanassopoulos:1997er}.  As
a model of a DAR source, we use the DAE$\delta$ALUS design
\cite{Alonso:2010fs}.

The DAE$\delta$ALUS accelerators are cyclotrons
\cite{Alonso:2010yz,Calabretta:2010mh, Calanna:2011gm}, an ideal
low-cost source for low energy (800 MeV) protons. The one caveat to
the use of a cyclotron as the accelerator is that the bunch spacing is
typically a few tens of nanoseconds ({\it e.g.} one DAE$\delta$ALUS
design operates at 66 MHz \cite{Calanna:2011gm}), hence much smaller
than the muon lifetime.  As a result these machines effectively
operate as a continuous source (often called a ``CW'' source).  It is
straightforward to run this CW beam for milliseconds and then turn the
beam off; however, implementing shorter spills is
costly. DAE$\delta$ALUS designs typically run the beam for $\sim$100
ms periods.  The required long spill precludes use of beam-timing to
identify flavors, as has been suggested for DAR studies at spallation
sources that employ expensive linacs and buncher rings
\cite{Garvey:2005pn,Mills:2009zz,Ray:2008ea}.

The DAE$\delta$ALUS accelerators are being designed to produce more
than 1 MW average power.  The instantaneous power is a product of the
energy and current of the protons on target, while the average power
also accounts for the duty factor.  The DAE$\delta$ALUS power needs
are driven by the multi-kilometer distances required for the
$CP$-violation search.  The extracted beam is defocussed and the
carbon target has a tapered entrance to spread the beam, to allow high
instantaneous power on target.  In order to cycle between sites and to
allow for beam off running, the DAE$\delta$ALUS experiment employs
10\% to 20\% duty factors.  This leads to high instantaneous power
requirements that are not needed in the SBL experiments we describe
here.

As will be shown in \Sec~\ref{Results}, an average beam power of 100
kW is sufficient to achieve most goals.  We propose to allow some
beam-off periods so that cosmogenic backgrounds can be measured and
subtracted. A reasonable run-plan could have a 50\% duty factor with
a 200 kW instantaneous power on the target, delivered with the same
time structure as planned for DAE$\delta$ALUS.
Table~\ref{tab:beamparams} provides an overview of the assumptions
used for the calculations in \Sec~\ref{Results}, unless otherwise
noted. The low power requirement, with similar time structure to
DAE$\delta$ALUS, would make these accelerators ideal as prototypes for
the high-power DAE$\delta$ALUS machines.

The position of the neutrino production point depends on the proton
interaction length in the target material, the pion stopping length in
the target, and shielding and tapering of the target entrance to
spread the beam power.  We will assume a $\pm 25$ cm smearing for the
neutrino production point in this study.  We take this to be flat in
$L$, which is an overestimate of the smearing.  We also assume that,
on average, neutrinos are produced 20 m from the detector face to
allow for accelerator shielding.  We assume DAE$\delta$ALUS-level
$\bar \nu_e$ contamination, which is at $4\times 10^{-4}$ of the
$\nu_e$ in the beam.

While the ratio of flavors and the energy dependence of the flux are
well understood in a DAR flux, the overall normalization is not well
predicted.  The dominant error comes mainly from the uncertainty on the pion
production per proton on target~\cite{Burman:1996gt,Allen:1989dt,Burman:1989dq}. 
We assume a 10\% correlated normalization error on
all flavors, which is slightly more conservative than in~\cite{Burman:1996gt,Allen:1989dt,Burman:1989dq}.  
We also assume a 20\% error on the $\pi^-$ DIF background~\cite{Athanassopoulos:1997er}.

\section{(3+n) sterile neutrino oscillation hypotheses}

If one assumes CPT invariance and no matter effects, then the probability
for a neutrino produced with flavor $\alpha$ and energy $E$, to be detected 
as a neutrino of flavor $\beta$ after traveling a distance $L$, 
is~\cite{Barger:1999hi,Kayser:2002qs}: 
%%%%%%%%%%
\begin{eqnarray}
\label{eq:oscprob}
P(\nu_{\alpha}\to\nu_{\beta})= & \delta_{\alpha\beta}-4\sum_{i>j}\mathcal{R}(U^{\ast}_{\alpha i}
 U_{\beta i}U_{\alpha j}U^{\ast}_{\beta j})\sin^2x_{ij}+\nonumber \\
 & 2\sum_{i>j}\mathcal{I}(U^{\ast}_{\alpha i}U_{\beta i}U_{\alpha j}U^{\ast}_{\beta j})\sin2x_{ij}
\end{eqnarray}
%%%%%%%%%%%
where \begin{math}\mathcal{R}\end{math}
and \begin{math}\mathcal{I}\end{math} denote the real and imaginary
parts of the product of mixing matrix elements, respectively.
In Eq.~\ref{eq:oscprob}, $\alpha,\beta\equiv e,\mu ,\tau$, or $s$,
($s$ being the sterile flavor); $i,j=1,\ldots ,3+n$ ($n$ being the
number of sterile neutrino species).  The neutrino mass splitting is
given by $\Delta m^2_{ij}\equiv m^2_i-m^2_j$ in $\hbox{eV}^2$.  The
$L/E$ dependence associated with oscillations appears within the term
$x_{ij}\equiv \Delta m^2_{ij}L/4E$, where $L$ is in m and $E$ is in
MeV.  The $U$ represent the elements of the mixing matrix that
connects the mass to the flavor eigenstates. For anti-neutrinos, the
oscillation probability follows Eq.~\ref{eq:oscprob} with the
replacement of $U$ with its complex-conjugate matrix. Therefore, if the 
elements of the mixing matrix are not real, neutrino and anti-neutrino oscillation
probabilities are not identical. Imaginary parameters enter the 
mixing matrix through $CP$-phases, which, in turn, lead to 
differences in the neutrino versus anti-neutrino oscillation probability.

In a (3+2) model, two sterile neutrinos are added at the eV scale with the three active 
neutrinos. In the SBL approximation where the largest mass-splittings dominate
and $\Delta m^2_{21} \approx \Delta m^2_{31} \approx 0$, 
following Eq.~\ref{eq:oscprob}, the $\bar\nu_{\mu} \to \bar\nu_e$ 
appearance oscillation probability is given by 
%%%%%
\begin{eqnarray}
\label{eq:threeplustwo_2}
P(\bar\nu_{\mu} \to \bar\nu_{e}) &=& 4|U_{e 4}|^2|U_{\mu 4}|^2\sin^2 x_{41} \nonumber \\
&+& 4|U_{e 5}|^2|U_{\mu 5}|^2\sin^2 x_{51} \nonumber \\
&+& 8|U_{e 4}U_{\mu 4}U_{e 5}U_{\mu 5}|\sin x_{41}\sin x_{51}\cos (x_{54}+\delta)  
\end{eqnarray}
%%%%%%
where $\delta \equiv arg(U_{e4}^*U_{\mu 4}U_{e5}U_{\mu 5}^*)$ is the $CP$-phase. 
We use this expression to estimate the signal event rates for appearance studies (see \Sec~5) 
in (3+2) models for the best fit parameter values as given in Table~\ref{tab:global-3+2}.   
We can see that the probability in Eq.~\ref{eq:threeplustwo_2} depends on two independent
mass splittings ($\Delta m^2_{41}$, $\Delta m^2_{51}$) and on the combinations
$|U_{e 4}U_{\mu 4}|$ and $|U_{e 5}U_{\mu 5}|$. Therefore, including $\delta$, the total number of 
independent parameters in this channel is 5.
  
In a (3+1) model with only one sterile neutrino, Eq.~\ref{eq:threeplustwo_2}
takes the form
%%%%%%%%%
\begin{eqnarray}
\label{eq:threeplusone_app}
P(\bar\nu_{\mu} \to \bar\nu_{e}) = 4|U_{e 4}|^2|U_{\mu 4}|^2\sin^2 x_{41}
\equiv \sin^22\theta_{\mu e}\sin^2 x_{41}
\end{eqnarray}
%%%%%%%%%
where $\sin^22\theta_{\mu e}$ is the effective mixing angle. One can see that $CP$-phase does not 
appear in the (3+1) case. 

Again following Eq.~\ref{eq:oscprob} with the SBL approximation, the $\nu_e \to \nu_e$ disappearance oscillation 
probability in a (3+2) model can be written as
%%%%%%%%%%
\begin{eqnarray}
\label{eq:threeplustwo_1}
P(\nu_e \to \nu_e) &=& 1-4(1-|U_{e 4}|^2-|U_{e 5}|^2)(|U_{e 4}|^2\sin^2 x_{41}+|U_{e 5}|^2\sin^2 x_{51})  \nonumber\\
&-& 4|U_{e 4}|^2|U_{e 5}|^2\sin^2 x_{54},
\end{eqnarray}
%%%%%%%%%%
which we use to simulate the survived event rates in (3+2) models. In this channel, we have 4 independent parameters.

In a (3+1) model, Eq.~\ref{eq:threeplustwo_1} simplifies to
%%%%%%%%%%%
\begin{eqnarray}
\label{eq:threeplusone_disapp}
P(\nu_{e} \to \nu_{e}) = 1- 4|U_{e4}|^2(1-|U_{e4}|^2) \sin^2 x_{41}
\equiv 1-\sin^2 2\theta_{ee} \sin^2 x_{41},
\end{eqnarray}
%%%%%%%%%%%
where $\sin^2 2\theta_{ee}$ is the defining mixing angle for SBL electron neutrino disappearance.
 
%%%%%%%%%%%%%%%%%%%%%%%%%%%
\begin{table}[t]
\begin{center}
  \begin{tabular}{ccccccccc}
\hline
  & $\Delta m^2_{41}$ & $|U_{e4}|$ & $|U_{\mu 4}|$ &
    $\Delta m^2_{51}$ & $|U_{e5}|$ & $|U_{\mu 5}|$ & $\delta / \pi$ \\
\hline
 A : Ref. \cite{Kopp:2011qd} & 0.47 & 0.128 & 0.165 & 0.87 & 0.138 & 0.148 & 1.64 \\
 B : Ref. \cite{Karagiorgi:2009nb} & 0.39 & 0.40  & 0.20  & 1.10 & 0.21  & 0.14  & 1.1  \\   
\hline
  \end{tabular}
\caption{The 1st row depicts the parameter values at the global best fit points for 3+2 model as described in~\cite{Kopp:2011qd}. The 2nd row shows the 3+2 best fit values from~\cite{Karagiorgi:2009nb} which has been derived using all the data sets of SBL appearance experiments. Here mass splittings are shown in eV$^2$.}
\label{tab:global-3+2}
\end{center}
\end{table}
%%%%%%%%%%%%%%%%%%%%%%%%%%%%

\section{Liquid Scintillator Detectors}

Scintillating--oil detectors use well understood technologies, but the
planned next steps involve significant increases in scale. Segmented
scintillator detectors have been traditionally used for GeV-scale
experiments. The largest pure-scintillator, finely-segmented example is
the 170t ton BNL 734 experiment \cite{Ahrens:1986ika}. On the other
hand, KamLAND~\cite{:2008ee} and Borexino~\cite{Arpesella:2008mt} have
demonstrated ton-scale unsegmented liquid scintillator detectors, with
the latter demonstrating a very low level of background, allowing
low-energy ($\sim$ 1 MeV) neutrino studies with low systematics.

The next generation of these detectors are more than an order of
magnitude larger than those described above.  NO$\nu$A, which is under
construction at Ash River, Minnesota, US, is a segmented detector.
This detector will come on-line in 2013 \cite{Feldman:2008zza}. LENA,
which is an unsegmented, very-low background detector, is proposed for
the LAGUNA \cite{Angus:2010sz} project in Europe.  Because site-selection for LAGUNA 
is only just underway, this detector will come online later than
NO$\nu$A, around 2020.

Both NO$\nu$A and LENA are intended for long baseline oscillation
studies.  The SBL running does not conflict with the long
baseline running for several reasons.  First, the long baseline
neutrino beam has energies $>100$ MeV, producing events that are
easy to separate from the lower energy DAR events.  Second, the long baseline
beam timing allows one to 
gate out the DAR beam for a few milliseconds around the
the long-baseline few-microsecond spill.

%%%%%%%%%%%%%%
\subsection{DAR beam interactions in liquid scintillator oil}
%%%%%%%%%%%%%%

%%%%%%%%%%%%%%%%%%%%%%%%%%%%%%%%%%%%%%%%%%%%%%%%%%%%%%%%
\begin{figure}[tp]
\centering
\includegraphics[width=0.5\textwidth]{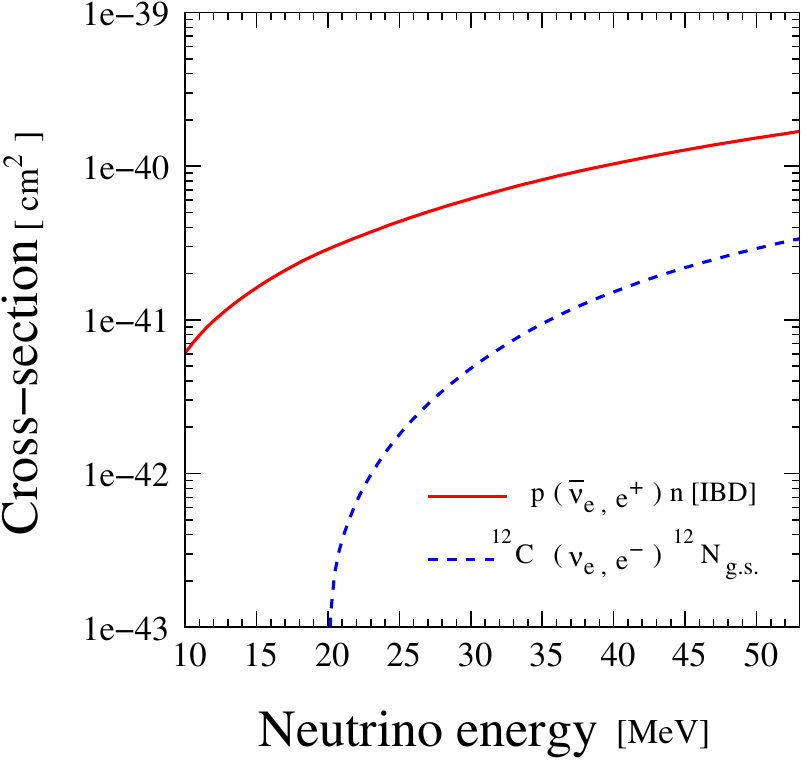}
\mycaption{\label{fig:flux-crosssection}  Cross-sections of all the flavor selective processes used in this study.}
\end{figure}
%%%%%%%%%%%%%%%%%%%%%%%%%%%%%%%%%%%%%%%%%%%%%%%%%%%%%%%%%

As discussed above, a DAR beam consists of $\nu_\mu$, $\bar\nu_\mu$, $\nu_e$ with a small admixture 
of $\bar \nu_e$ with energies ranging up to 52.8 MeV. Because of the low beam energy, a muon cannot
be produced in a CC interaction. Therefore, detectors with oil-based targets (CH$_n$) are limited to 
observe CC interactions involving the electron flavor scattering. Inverse $\beta$-decay, (IBD), 
$\bar \nu_e + {\rm p} \to e^+  +  {\rm n}$ ~\cite{Vogel:1999zy} is the golden channel for these 
detectors to detect the appearance of $\bar\nu_e$. This process has a very low kinematic threshold 
of 1.81 MeV and provides a very useful delayed coincidence tag between the prompt positron and the
delayed neutron capture by a proton, ${\rm n + p} \to {\rm d} + \gamma$ (2.2 MeV) after an average
time of 250 $\mu$s.

Another CC reaction that we consider is 
$\nu_e + ^{12}$C $\to e^- + ^{12}$N$_{\rm g.s.}$~\cite{Auerbach:2001hz} with a relatively high 
kinematic threshold of 17.33 MeV. This process is used to perform the disappearance searches 
using $\nu_e$. In this cross-section, we only consider the contribution coming from the transition
to the $^{12}$N ground state, which can be identified by the detection of the prompt electron, 
followed within a 60 ms window by the positron from the beta decay of the $^{12}$N$_{\rm g.s.}$ with a mean 
lifetime of 15.9 ms. The cross section for this exclusive reaction $^{12}$C($\nu_e$, e$^{-}$)$^{12}$N$_{\rm g.s.}$
is well measured compared to the inclusive cross-section for transitions to excited states of $^{12}$N
and has an error of 5 to 10\%~\cite{Auerbach:2001hz}. Probably the contribution from transitions 
to excited states of $^{12}$N can also be used and would increase the event rate by about
30 to 40\%. The excited state cross-sections are less well known but since the power of the measurements 
presented here mainly comes from an $L/E$ shape analysis, these uncertainties are not very important.  

Fig.~\ref{fig:flux-crosssection} shows the cross sections both of 
the CC flavor selective processes used 
in this study. The cross section for  
$\bar\nu_e + ^{12}$C $\to e^+ +^{12}$B  compared to that for IBD scattering
is sufficiently low to be neglected from these studies. We also do not consider the neutrino-electron scattering 
interaction because the rates are two orders of magnitude lower than the rate of $^{12}$C scattering. 

%%%%%%%%%%%%%%%%%%%%%%%%%%%%%%%%%%%%%%%%%

\begin{table}[t]

\begin{center}
\begin{tabular}{||c||c||c||} \hline\hline
\multicolumn{1}{||c||}{{\rule[0mm]{0mm}{6mm}{Detector}}}
& \multicolumn{1}{|c||}{\rule[-3mm]{0mm}{6mm}{NO$\nu$A Far}}
& \multicolumn{1}{|c||}{\multirow{2}{*}{\rule[-3mm]{0mm}{6mm}{LENA}}}
\cr
Characteristics & Detector &
\cr
\hline\hline
Shape & Rectangular & Cylindrical \cr
\hline
Fiducial Mass & $14\,\mathrm{kt}$ & $(5 - 50)\,\mathrm{kt}$  \cr
\hline
Overburden & 3 m earth-equivalent & 1450 m of rock/4060 mwe  \cr
 & &   @ Pyh{\"a}salmi \cr
\hline
Solvent&  ${\rm C}{\rm H}_{2}$ & LAB (${\rm C}_{18}{\rm H}_{30}$)  \cr
\hline
Neutrino Energy & \multirow{2}{*}{$38\,\mathrm{MeV}$ (Dis)} & $20\,\mathrm{MeV}$ (App)  \cr
Threshold & & $33\,\mathrm{MeV}$ (Dis)  \cr
\hline
\multirow{2}{*}{Detection Efficiency} & \multirow{2}{*}{50\% (Dis)} & 90\% (App)  \cr
& & 80\% (Dis) \cr
\hline
Energy Resolution, $\sigma({\rm E})$ & \multirow{2}{*}{15\%$\sqrt{\rm E/MeV}$~\cite{messier}} & \multirow{2}{*}{10\%$\sqrt{\rm E/MeV}$~\cite{Wurm:2007cy}} \cr
in units of $\mathrm{MeV}$ & & \cr
\hline
\multirow{2}{*}{Signal error (syst.)} & \multirow{2}{*}{20\% (Dis)} & 10\% (App)  \cr
& & 15\% (Dis) \cr
\hline
\multirow{2}{*}{Background error (syst.)} & \multirow{2}{*}{5\% [Non-beam] (Dis)} & 20\% [Intrinsic $\bar\nu_e$, $4 \times 10^{-4}$] (App)  \cr
& & -- \cr
\hline\hline
\end{tabular}
\caption{\label{tab:detector} Detector characteristics used in the simulations.
``App'' applies to the appearance analysis. ``Dis'' applies to the disappearance analysis.}

\end{center}

\end{table}

%%%%%%%%%%%%%
\subsection{A Finely-Segmented Detector: NO$\nu$A}
\label{nova}
%%%%%%%%%%%%%

NO$\nu$A~\cite{Ayres:2007tu} is a segmented, scintillator detector constructed
for long baseline neutrino oscillation studies around 2 GeV, which is 40
times higher in energy than the DAR events.  Because this detector is
not designed for $\sim$~50 MeV events, it is not ideal for the studies
we present here.  Nevertheless, we show that NO$\nu$A
has the capability to produce significant disappearance results with a few
years of DAR beam running.

NO$\nu$A consists of PVC cells filled with scintillator oil, assembled
into a $15.7 \times 15.7 \times 67$ m$^3$ detector.  The scintillator
is mineral oil base doped with 4.1\% pseudocumene
[1,2,4-Trimethybenzene].  The PVC structure (CH$_2-$CHCl), which is
dead region, represents 30\% of the 14 kt fiducial mass. For
simplicity, we assume that the target as entirely CH$_2$ for this
analysis.  Individual cells are 3.9 cm high $\times$ 6.0 cm along the
beam direction $\times$ 15.5 m long transversely and are glued in $X$
and $Y$ planes. The orientation of the planes alternates throughout
the detector.

Light produced in the scintillator is absorbed and re-emitted with
wavelength shifting (to 550 nm) optical fibers.  These run the length
of the PVC cell and double back to readout APDs.  The APDs have a rather
high noise rate but will be cooled to reduce the noise to be below the
cosmic-ray muon rate. The attenuation
length of the fiber is roughly the same as the 15 m length of the cells.
Triggering requires at least one 1 hit in adjacent $X$ and $Y$ planes.
The detector is below ground level with overburden shielding of 3m
of earth.  This modest shielding leads to a 20 kHz rate of
through-going and 1 kHz of stopping muons
\cite{messier} in the detector.

The NO$\nu$A setup is ideal for the proposed cyclotron-dump design.
The cyclotron can be located on the surface, allowing easy maintenance,
while the target/dump can be located in an open area of the detector
hall which is used for staging and not needed after the detector is
constructed.  The incoming beam will impinge on a target
embedded in the dump. Targeting the beam perpendicular to the 
detector direction
is an extra precaution against DIF backgrounds.
In addition, the $\sim$~50 m length of the detector is
appropriate for neutrino oscillation measurements with 20 to 50 MeV neutrinos
probing the $\Delta m^2 \sim 1$ eV$^2$ region.

We expect that the shielding around the dump will lead to a very low
fast neutron background.  Because the detector staging area is quite
large, it will be possible to add up to 10 m of extra shielding, if
required.  Beyond this, the detector is self-shielding.  Therefore we
do not expect a significant fast neutron background from the
cyclotrons.  It is for this reason that we can place the face of the
detector at 20 m from the center of the target.  A summary of the
assumptions we make about the NO$\nu$A detector 
in the analyses is given in Table~\ref{tab:detector}.

The fine segmentation of the detector allows determination of the
vertex to a few cm.  As a result, the smearing in the measured $L$ 
value will be dominated by
the uncertainty in the neutrino production point for which we use
a $\pm 25$ cm flat distribution.

The disappearance study makes use of 
$\nu_e + ^{12}$C $\to e^- + ^{12}$N$_{\rm g.s.}$ interaction with 
a visible reconstruction
energy threshold of 20 MeV which corresponds to a neutrino energy threshold of 38 MeV.  
This threshold significantly reduces the accepted $\nu_e$ flux for the measurements,
as seen in Fig.~\ref{fig:flux} but still gives about 34k events per year (see \Sec~\ref{Results}).
We assume 50\% efficiency for this process.

The $\nu_e$ scatters will be reconstructed along the 67 m length of
the detector, allowing the oscillation wave to be fit as a function of
$L/E$. Since the target nucleus is much heavier than the
outgoing electron, the incident neutrino energy is simply related to
the the visible outgoing electron energy independent of the scattering angle.
The electron energy resolution is listed in Table~\ref{tab:detector}.

The largest background source will be from the $10^{10}$ Michel
electrons/year produced by stopped muon decay.  These are produced
with nearly the same energy dependence as the electrons from
CC $\nu_e$ scatters.  
The Michel electron events can be identified and vetoed by tracking the parent muon.
The entire target can be used to identify muons before a
candidate $\nu_e$ scattering event, which may require that a
substantial part of the upper region of the target be used for this veto.
In the analysis presented in the following section, we consider a range of veto
capabilities giving from 10,000 to 50,000 
total un-vetoed Michel background events per year.  We
also show the no-background case for reference.  We assume that the 50\% beam-off running 
can measure this background with a 5\% error in each bin.

In principle, the  NO$\nu$A detector can also tag $\bar \nu_e + {\rm p} \to e^+
+ {\rm n}$ events through the coincidence signal of the positron light
followed by light from the 2.2 MeV photon emitted in the neutron
capture on H.  In practice, observation of the the neutron-capture signal in NO$\nu$A will be quite difficult. 
First, the signal from the 2.2 MeV $\gamma$ is diffuse, being produced by electrons from
multiple Compton scatters. Because the NO$\nu$A detector has 30\% dead
material, there is a high probability that a substantial fraction of the 2.2 MeV
energy deposit will be unobserved.  Second, the attenuation length of
the wavelength shifting fibers is on the scale of the 15 m length of
the scintillator tubes.  Thus, the detector is highly inefficient for
very low light deposits. Third, the false coincidences with comic muon
produced background may be very high in a surface detector.
These are complex analysis issues that would require further study to quantify and
we therefore choose not to present the appearance capability of NO$\nu$A in this paper.

%%%%%%%%%%%%%%%%%%%%%%%
\subsection{An Unsegmented Detector: LENA}
\label{sec:lena}
%%%%%%%%%%%%%%%%%%%%%%%

The LENA (Low Energy Neutrino Astronomy) detector~\cite{Wurm:2011zn}
is optimized for low energy neutrino detection.  The experiment is specifically designed
for state-of-the-art measurements of supernova neutrinos (bursts and
diffuse), which are in the same energy range as DAR flux, and solar
neutrinos, geoneutrinos, and reactor neutrinos, all of which are an
order of magnitude lower in energy.  As a result, this detector is
substantially more powerful than NO$\nu$A for the DAR-based
SBL program we consider here.

The low energy neutrino studies require substantial shielding of at least
4000 mwe. The search for a possible underground site for LENA has been performed
in the framework of the LAGUNA design study~\cite{Angus:2010sz}.
Two possible choices are Pyh{\"a}salmi and Fr{\'e}jus.
Both the sites have rock shielding $>$ 4000 mwe resulting a very low
rates of through-going muons, at the level of
$\sim 5\times10^{-5}$ Hz/m$^2$~\cite{Wurm:2011zn}.
The stopping muon background rate is negligible.  The rate of atmospheric electron
neutrino interactions in the DAR energy range will also be negligible.
Nevertheless, we assume that there will be 50\% beam-off running.  
This allows cross-checks
of the backgrounds as well as allowing the simultaneous data-taking
for the supernova and SBL studies.

The LENA detector is planned to be 100 m in length and 30 m in diameter.
If a fiducial volume of 13.6 m radius is chosen, then the detector will contain
50 kt of liquid scintillator while the outside region will be
filled with water to act as a veto for muons and shield for neutrons.
The detector will have 30\% coverage from a combination of direct
photocathode coverage and light concentrators.

In the studies below, we also consider several smaller fiducial
volumes.  None are in the present LENA plans.  However, a 5 to 10 kt
prototype may be attractive along the path to LENA and pressure at the
base of the detector on the PMTs may lead preferance for a 25 kt
design.  Therefore, we also include these possibilities in our
discussion.

Both CC $^{12}$C events for disappearance and IBD scattering events
for appearance can be used in the
analysis. For LENA, we consider LAB solvent (${\rm C}_{18}{\rm H}_{30}$) with 
$6.6 \times 10^{28}/{\rm m}^{3}$ free protons and
$4.0 \times 10^{28}/{\rm m}^{3}$ carbon nuclei in a 50 kt detector. 
Demonstration that oil can be purified well below the levels required for the SBL 
physics come from the KamLAND~\cite{:2008ee} and
Borexino~\cite{Arpesella:2008mt} experiments. The specific assumptions about the 
LENA detector relevant to this study appear in Table~\ref{tab:detector}.

The arrangement of the cyclotron and target/dump will be designed
after the location of the detector is established.  Tunnel access
where the neutrino source can be moved into the detector region is most attractive.
We assume that the neutrino source will be located on the long-axis of
the cylinder and will be 20 m away from the cylinder surface.

LENA can explore both appearance and disappearance oscillations.
The signals are the same as discussed for NO$\nu$A: $\bar\nu_e + {\rm p} \to 
e^{+} + {\rm n}$ and $\nu_e + ^{12}$C $\to e^- + ^{12}$N$_{\rm g.s.}$,
respectively. Because the detector is designed for excellent reconstruction of
geoneutrinos and astrophysical signals at 2 MeV, there is high
efficiency for reconstructing the 2.2 MeV neutron capture $\gamma$.
We consider 90\% detection efficiency for appearance studies
and 80\% efficiency for disappearance, although this is likely to be very conservative.
Vertex reconstruction can be expected to have a 5 cm uncertainty~\cite{Wurm:2011zn}.  
This is negligible on the scale of the $\pm$25 cm uncertainty from the neutrino source.

We assume that there is very little background in LENA.  The $>4000$ mwe
shielding reduces the cosmic ray background to a negligible level.
The next largest natural background is from atmospheric neutrinos.
These are at the $<$1\% level for the 100 kW source.  It should be
noted that the atmospheric muon neutrino background is much lower than
in a water Cerenkov detector.  This is because, in a scintillation
detector, all muons from charged current events are identified,
whereas in the water detector some are below Cerenkov threshold.  We
assume a $>10$ m wall of undisturbed rock (or equivalent steel or concrete
shielding) in the design that will reduce beam backgrounds from neutrons produced at the accelerator 
to a negligible level at the detector.  The additional water shielding further protects the detector.

Disappearance searches proceed in the same manner as NO$\nu$A.  The
signal is the variation of $\nu_e + ^{12}$C $\to e^- + ^{12}$N$_{\rm g.s.}$ across the length 
of the detector. For the oscillation sensitivity estimates, the simulated
data, taking into account the detector energy resolution and the neutrino source 
position smearing, are binned according to energy and 
position in 65 equally sized $L/E$ bins.  A visible energy
cut of 16 MeV is imposed for the events, which corresponds to a 33 MeV cut
on the incoming neutrino energy.  This visible energy cut will practically
eliminate backgrounds from atmospheric and reactor neutrinos as well as
environmental radioactivity.
Normalization is included but because these errors are large,
the fits are dominated by the $L/E$ shape dependence.

LENA is designed also to have high efficiency for 
$\bar\nu_e + {\rm p} \to e^{+} + {\rm n}$ (IBD) events at neutrino energies 
in the 10 to 50 MeV range identified by a delayed coincidence between the 
outgoing positron and capture of the neutron. This is the signal for diffuse supernova
neutrinos, and an important process for supernova burst events -- both of
which are key physics goals of LENA. Thus, LENA is specifically
designed for lower energies in contrast to NO$\nu$A.  Using the IBD events,
LENA can make a very precise search for $\bar\nu_e$ appearance by again
binning and fitting events as a function reconstructed $L/E$.  
For our studies, the visible energy threshold is set to 19 MeV corresponding to a neutrino energy 
cut of 20 MeV.  Again, this requirement will reduce the non-beam backgrounds to negligible level.

This puts LENA in a unique position to measure both appearance and
disappearance in the same detector through the characteristic change
in the oscillation wave with $L$.  That would be a very powerful
signal in support of sterile neutrino models and offers the best
opportunity to disentangle (3+1) from (3+2) oscillations.

%%%%%%%%%%%%%%%%
\section{Results \label{Results}}
%%%%%%%%%%%%%%%%

In this section, we first discuss SBL $\bar \nu_\mu \rightarrow \bar\nu_e$ appearance with LENA.  
This is key to confirming or refuting the
LSND~\cite{Athanassopoulos:1995iw,Athanassopoulos:1996jb,Athanassopoulos:1997er,Aguilar:2001ty}
and MiniBooNE~\cite{AguilarArevalo:2010wv} anti-neutrino results. Then we focus on SBL $\nu_e$ disappearance 
studies with LENA and NO$\nu$A, which are complementary searches to the reactor anomaly
$\bar\nu_e$ disappearance signal~\cite{Mention:2011rk,Mueller:2011nm}.

%%%%%%%%%%%%%%%%
\subsection{Appearance Mode}
%%%%%%%%%%%%%%%%

%%%%%%%%%%%%%%%%%%%%%%%%%%%%%%%%%%%%%%%%%%%%%%%%%%%
\begin{table}[t]
\begin{center}
  \begin{tabular}{||c||c||c||c||c||c||c||} \hline \hline
    Fiducial Mass & Radius & Length & Signal      & Signal      & Intrinsic $\bar\nu_e$ \\
                  &        &        & (A : Ref. \cite{Kopp:2011qd})  &  (B : Ref. \cite{Karagiorgi:2009nb}) &  Background\\ 
    \hline
    50 kt & 13.58 m & 100 m & 12985 & 32646 & 1450 \\
    \hline
    25 kt & 10.78 m & 79.37 m & 7787 & 18356  & 875 \\
    \hline
    10 kt & 7.94 m & 58.48 m & 3753 & 7964 & 443 \\
    \hline
    5 kt & 6.3 m & 46.42 m & 2080 & 4044 & 261 \\
    \hline \hline
\end{tabular}
\caption{\label{tab:ls-app-events} The expected number of signal and
intrinsic beam background events in 5 to 50 kt LENA detector. 
While varying the fiducial mass of the detector, we have kept the density and the aspect ratio (length/diameter) 
same in all the cases. In column 4 and 5, the signal events have been computed using the two different sets of parameter 
values in (3+2) model as given in Table~\ref{tab:global-3+2}. 
Here we have used total $4 \times 10^{21}$ $\bar\nu_{\mu}$.
The intrinsic $\bar\nu_e$ beam contamination is $4\times10^{-4}$.}
\end{center}
\end{table}
%%%%%%%%%%%%%%%%%%%%%%%%%%%%%%%%%%%

%%%%%%%%%%%%%%%%%%%
\begin{figure}[t]
\centering
\includegraphics[width=0.5\textwidth]{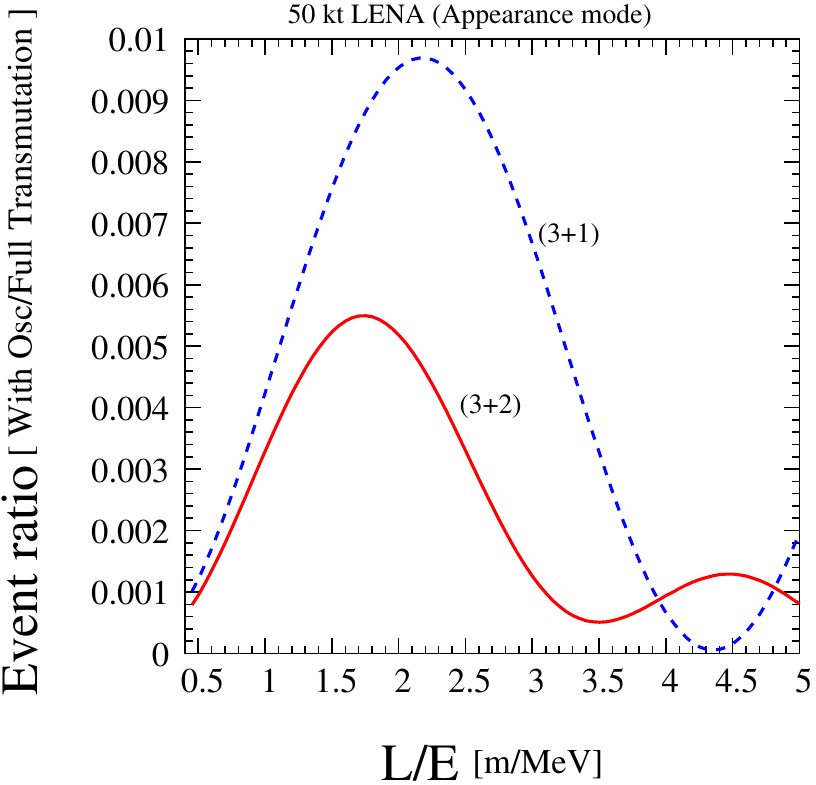}
\mycaption{\label{fig:ls-app-spec} The ratio of signal events estimated for a given set of oscillation 
parameter values and the events which have been computed assuming a hypothetical case where 
the $\bar\nu_{\mu} \to \bar\nu_e$ oscillation probability is one is shown as a function of reconstructed 
$L/E$ in 50 kt LENA. Here we have used total 
$4 \times 10^{21}$ $\bar\nu_{\mu}$. The solid red line is for the parameter values      
in (3+2) model as given in the upper row of Table~\ref{tab:global-3+2}. The blue dashed line is computed 
for (3+1) best fit values: $\Delta m^2_{41}$ = 0.57 and $\ts_{\mu e}$ = 0.0097 as given in~\cite{Karagiorgi:2009nb}.
See text for more details.}
\end{figure}
%%%%%%%%%%%%%%%%%%%%%%%%%%%%%%

The LENA appearance search uses the IBD signal. 
Following Ref.~\cite{Wurm:2011zn}, we consider a cylindrical unsegmented
liquid-scintillator detector of 50 kt fiducial mass with 100 m in length and 13.58 m in radius as a 
reference choice. We also study the impact of smaller-sized LENA type detectors with fiducial 
masses of 25, 10 and 5 kt. In all cases, we keep the fractional photodetector coverage 
and aspect ratio (length/diameter) same as we consider for 50 kt.
In addition, while varying the fiducial mass of the detector, we keep the detector characteristics unchanged
as described in Table~\ref{tab:detector}. The main source of background is the intrinsic 
$\bar\nu_e$ beam contamination from $\pi^{-}$ decays (see Table~\ref{tab:beamparams}),
because cosmogenic backgrounds are very low (see \Sec~\ref{sec:lena}).
The neutrino energy threshold for this analysis is $20\,\mathrm{MeV}$ which renders 
potential backgrounds coming from supernova and radioactive decay negligible. 
We present all our results assuming a flat resolution on the neutrino
length distribution of $\pm$25 cm dominated by the uncertainty in the neutrino production point.

In Table~\ref{tab:ls-app-events} we present the signal and intrinsic $\bar\nu_e$ background event
rates in 5 to 50 kt LENA type detector including the source and the detector characteristics as shown in
Table~\ref{tab:beamparams} and~\ref{tab:detector} respectively. In column 4 and 5, the signal events
have been estimated for two different sets of parameter values in (3+2) model as given in 
Table~\ref{tab:global-3+2}. Here we consider a total neutrino flux of
$4 \times 10^{21}$ $\bar\nu_{\mu}$ from DAR beam. One should note that the number of signal
events does not increase linearly with the fiducial mass because
of the geometry of the detector and the isotropic nature of the flux. While increasing the fiducial mass from
5 to 50 kt, the length of the detector increases by a factor of 2.15 and since the source is located on the axis
of the cylinder and 20 m away from the beginning of the detector, a very small amount of
neutrinos reach the ends of the detector due to the 1/${L}^2$ suppression from the isotropic source.

With a source-to-detector-face distance of 20 m, the accessible L range in 50 kt  detector is 20 - 120 m and with an energy
range of 20 - 52.8 MeV in DAR beam, the $L/E$-dependence of the oscillation pattern can be well observed inside the
detector. This is illustrated in Fig.~\ref{fig:ls-app-spec}, where in each reconstructed $L/E$ bin, we show the ratio
of signal events for a given set of oscillation parameter values to the number of events
computed assuming a hypothetical case where the $\bar\nu_{\mu} \to \bar\nu_e$ oscillation probability is one.
This has been done just to demonstrate the oscillation pattern inside the detector as a function of reconstructed
$L/E$ for a particular set of parameter values. The red solid line corresponds to the parameter values
in the (3+2) model as given in the upper row of Table~\ref{tab:global-3+2}, denoted by A.
The blue dashed line has been drawn for the (3+1) best fit values ($\Delta m^2_{41}$ = 0.57 eV$^2$ and $\ts_{\mu e}$ = 0.0097)
obtained using LSND, MiniBooNE anti-neutrino and KARMEN~\cite{Armbruster:2002mp} data sets
in~\cite{Karagiorgi:2009nb}. As one can see from Fig.~\ref{fig:ls-app-spec}, the demonstration of the
oscillation wave is dramatic in the long LENA detector and can provide a powerful handle to discriminate
between (3+1) and (3+2) schemes.

For our statistical analysis, we bin our signal and background events into 94 equally sized $L/E$ bins and
consider uncorrelated normalization errors on the signal and intrinsic $\bar\nu_e$ background  of
10\% and 20\% respectively. These uncertainties are fully correlated between the  $L/E$ bins and
are included in the analysis using the pull-term method as described in {\it e.g.} Ref.~\cite{Huber:2002mx,Fogli:2002au}. 
For the fitting, we perform the usual $\chi^{2}$ analysis using a Poissonian likelihood function.

In Table~\ref{tab:ls-app-flux} we present the amount of $\bar\nu_{\mu}$ flux needed for a
5 to 50 kt LENA type detector to exclude the two different sets of oscillation parameter values
in the (3+2) scheme as given in Table~\ref{tab:global-3+2}. The numbers presented here for $5\,\sigma$ CL (5 dof) under
the assumption that we have only intrinsic $\bar\nu_e$ beam background. We can immediately see that
with a very modest flux and a small size LENA detector, we can check these test points at high significance. 
This also shows that the $L/E$ dependence is very important
for rejecting the background and therefore reducing the sensitivity to systematic errors.
The ability to observe the $L/E$ dependence is crucial if a signal is observed, since it will allow one to 
establish or refute oscillations as the underlying physics explanation.

%%%%%%%%%%%%%%%%%%%%%%%%%%%%%%%%%%%%%%%%%%%%%%%%%%%
\begin{table}[t]
\begin{center}
  \begin{tabular}{||c||c||c||c||} \hline \hline
    Fiducial Mass & Flux & Flux \\
                  &  (A : Ref. \cite{Kopp:2011qd})   &  (B : Ref. \cite{Karagiorgi:2009nb}) \\
    \hline
    50 kt & $0.912 \times 10^{19}$ & $0.302 \times 10^{19}$ \\
    \hline
    25 kt & $1.535 \times 10^{19}$ & $0.539 \times 10^{19}$ \\
    \hline
    10 kt & $3.235 \times 10^{19}$ & $1.27 \times 10^{19}$ \\
    \hline
    5 kt & $5.935 \times 10^{19}$ & $2.6 \times 10^{19}$ \\
    \hline \hline
\end{tabular}
\caption{\label{tab:ls-app-flux} It shows the amount of neutrino flux needed to exclude 
the two different sets of oscillation parameter values as given in Table~\ref{tab:global-3+2} 
at $5\,\sigma$ CL with 5 degrees of freedom ($\Delta\chi^2 = 37.09$) using 5 to 50 kt LENA 
in appearance mode.}
\end{center}
\end{table}
%%%%%%%%%%%%%%%%%%%%%%%%%%%%%%%%%%%%%%%%%%%%%%%%%%

\begin{figure}[tp]
\centering
\includegraphics[width=0.49\textwidth]{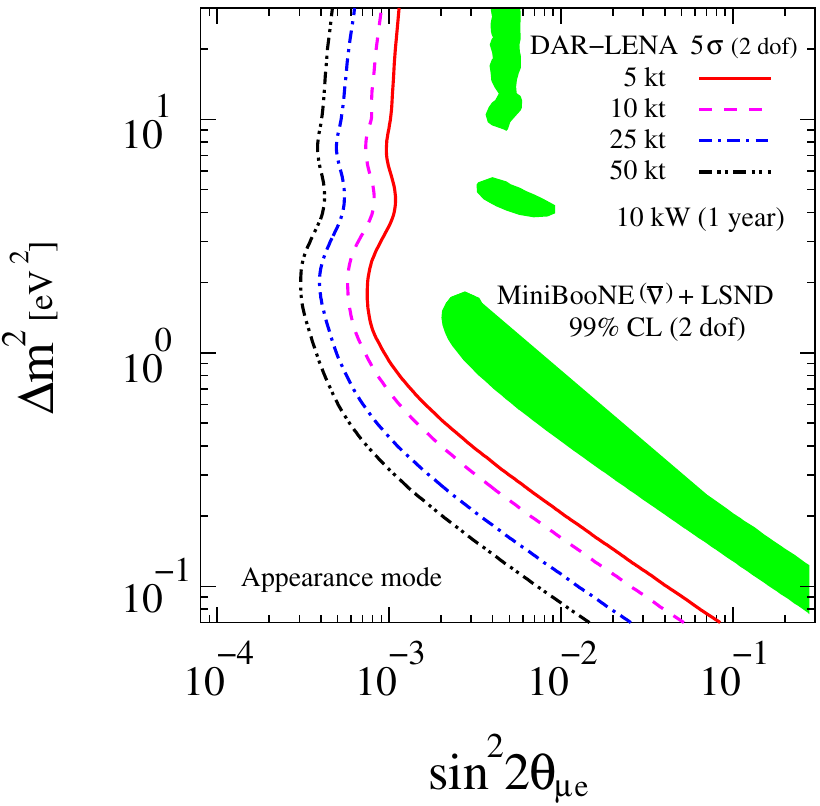}
\includegraphics[width=0.49\textwidth]{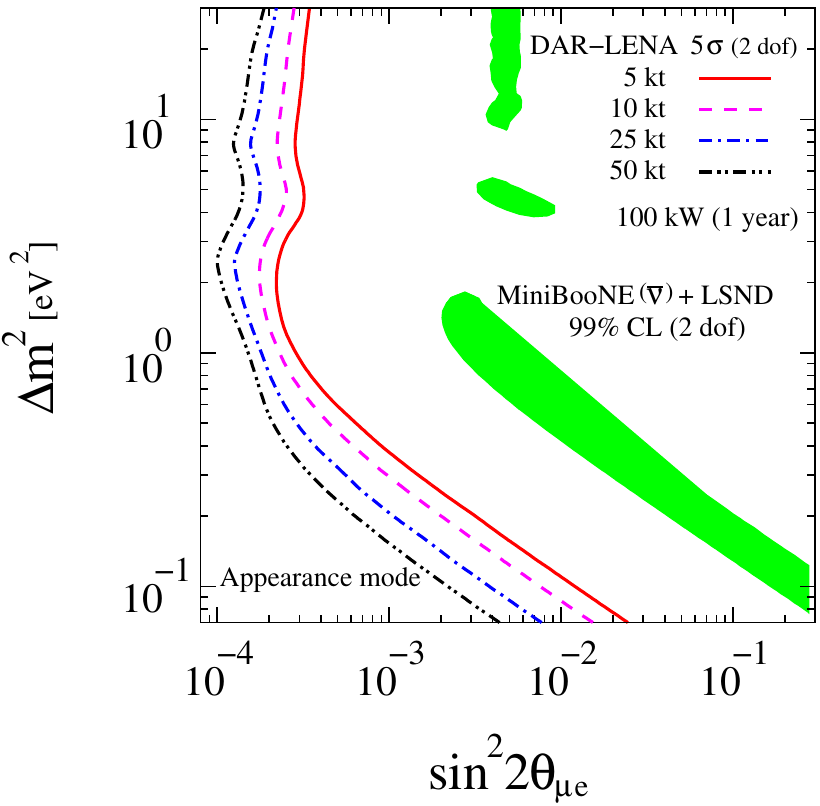}
\mycaption{\label{fig:ls-app-sens} Sensitivity limit of DAR-LENA setup to sterile neutrino oscillation in the (3+1) 
model at $5\,\sigma$ CL (2 dof) using appearance mode. The green/gray shaded area 
is the allowed region at 99\% CL (2 dof) from a combined analysis of the LSND and MiniBooNE anti-neutrino 
signals~\cite{Kopp:2011qd}. Left (right) panel shows the results for 10 (100) kW average power machine
which can deliver $4 \times 10^{20}$ ($4 \times 10^{21}$) $\bar\nu_{\mu}$ in one year.
In both the panels, results are shown for four different choices of the fiducial mass of the detector.}
\end{figure}

Fig.~\ref{fig:ls-app-sens} presents the sensitivity limit of the DAR-LENA
setup to sterile neutrinos in the (3+1)
model at 5$\,\sigma$ confidence level (2 dof) using the appearance mode.
We compare our results with the
allowed region at 99\% CL (2 dof) from a combined analysis of the LSND and MiniBooNE anti-neutrino
signals~\cite{Kopp:2011qd}. The results are presented for four different choices of the fiducial mass of the
detector. The left panel shows the sensitivity for
$4 \times 10^{20}$ $\bar\nu_{\mu}$ which can be achieved in one year with a small 10 kW average
power machine. The right panel exhibits the same for our reference choice of 100 kW machine.
It can be seen from the left panel that a 5 kt LENA type detector and a one year run at 10 kW
average power is more than sufficient to exclude the parameter space suggested by the
combined fit of LSND and MiniBooNE anti-neutrino data at 5$\,\sigma$ CL.
Note, that the sensitivity is limited by the magnitude of the beam background
and therefore does not improve linearly with the size of the detector.
In the right panel, the 50 kt detector has a reach up to $\ts_{\mu e}$ = 0.0001
at $\Delta m^2_{41}$ = 2 eV$^2$ with $4 \times 10^{21}$ $\bar\nu_{\mu}$.

%%%%%%%%%%%%%%%%%%%%%%%%%%%%%%%%%%%%%%%%%%%%%%%%%%%%%%%%
%%%%%%%%%%%%%%%%%%%%%%%%%%%%%%%%%%%%%%%%%%%%%%%%%%%%%%%%%
 
%%%%%%%%%%%%%%%%%%%
\subsection{Disappearance Mode}
%%%%%%%%%%%%%%%%%%%

%%%%%%%%%%%%%%%%%%%%%%%%%%%%%%%%%%%%%%%%%%%%%%%%%%%
\begin{table}[t]
\begin{center}
  \begin{tabular}{||c||c||c||c||c||c||c||} \hline \hline
    Fiducial Mass & Radius & Length & Evts w/ Osc & Evts w/ Osc & Evts, No Osc \\
              &        &        &  (A : Ref. \cite{Kopp:2011qd})   &  (B : Ref. \cite{Karagiorgi:2009nb})  &  \\
    \hline
    50 kt & 13.58 m & 100 m & 170191 & 139119 & 181672 \\
    \hline
    25 kt & 10.78 m & 79.37 m & 102726 & 85271 & 109590 \\
    \hline
    10 kt & 7.94 m & 58.48 m & 52105 & 43940 & 55439 \\
    \hline
    5 kt & 6.3 m & 46.42 m & 30874 & 26321 & 32735 \\
    \hline \hline
\end{tabular}
\caption{\label{tab:ls-disapp-events} The total number of CC $\nu_e$ scattering events on $^{12}$C in 
5 to 50 kt LENA. In column 4 and 5, the survived number of events have been computed using the two different 
sets of oscillation parameter values in (3+2) model as given in Table~\ref{tab:global-3+2}. The last column 
shows the total event rate without any oscillation. Here we have used total $4 \times 10^{21}$ $\nu_e$
with an energy threshold of 33 MeV and an efficiency of 80\%.}
\end{center}
\end{table}
%%%%%%%%%%%%%%%%%%%%%%%%%%%%%%%%%%%%%%%%%%%%%%%%%%

%%%%%%%%%%%%%%%%%%%%%%%%%%%%%%%%%%%%%%%%%%%%%%%%%%%
\begin{table}[t]
\begin{center}
  \begin{tabular}{||c||c||c||c||c||c||c||c||} \hline \hline
    Fiducial & Length & Breadth & Height & Evts w/ Osc & Evts w/ Osc & Evts, No Osc \\
    Mass &        &         &        &  (A : Ref. \cite{Kopp:2011qd})  &  (B : Ref. \cite{Karagiorgi:2009nb}) & \\
    \hline
    14 kt & 67 m & 15.7 m & 15.7 m & 32388 & 27407 & 34415 \\
    \hline \hline
\end{tabular}
\caption{\label{tab:nova-disapp-events} The expected number of CC $\nu_e$ scattering events on $^{12}$C 
in 14 kt  NO$\nu$A far detector. In column 5 and 6, the survived events have been computed 
using the two different sets of oscillation parameter values in (3+2) scheme as given in Table~\ref{tab:global-3+2}.
In the last column, we have the total event rate without any oscillation. Here we have used total 
$4 \times 10^{21}$ $\nu_e$ with an energy threshold of 38 MeV and an efficiency of 50\%.}
\end{center}
\end{table}
%%%%%%%%%%%%%%%%%%%%%%%%%%%%%%%%%%%%%%%%%%%%%%%%%%

\begin{figure}[tp]
\centering
\includegraphics[width=0.5\textwidth]{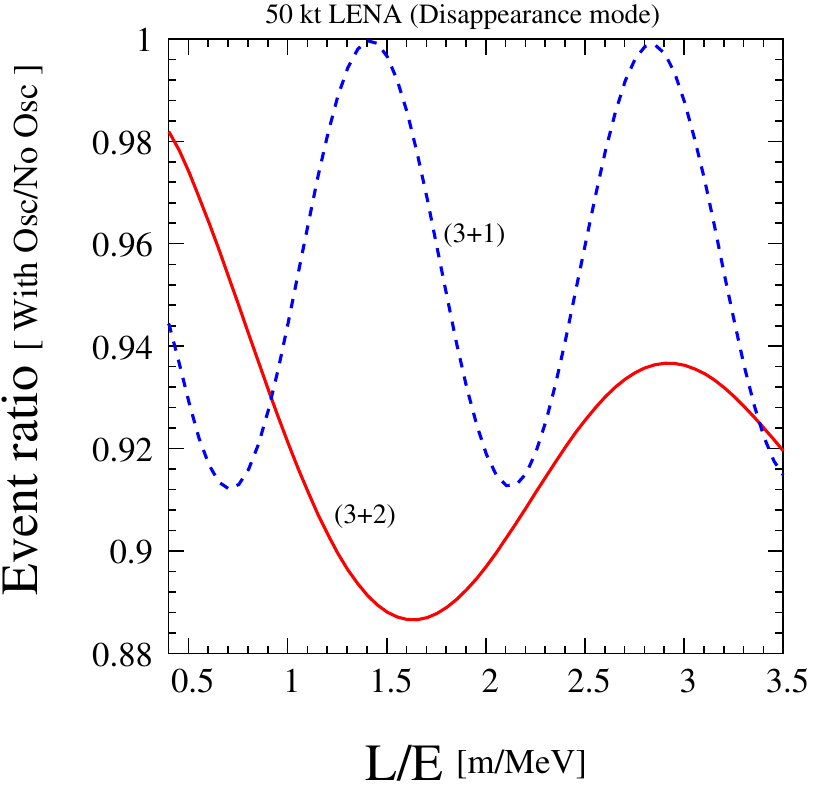}
\mycaption{\label{fig:ls-disapp-spec} This figure shows the ratio of events with and without sterile oscillation
as a function of the reconstructed $L/E$ in 50 kt LENA. Here we have used total $4 \times 10^{21}$ $\nu_e$. 
The solid red line is for the parameter values in (3+2) model as given in the upper row of Table~\ref{tab:global-3+2}. 
The blue dashed line is computed for (3+1) best fit values: $\Delta m^2_{41}$ = 1.78 and $\ts_{ee}$ = 0.089
as obtained from~\cite{Kopp:2011qd} using reactor anti-neutrino data with new predictions for the reactor flux.}
\end{figure}

Both LENA and NO$\nu$A offer the possibility to study oscillations to sterile neutrinos in the disappearance mode 
by using the CC reactions of $\nu_e$ on $^{12}$C. In Table~\ref{tab:ls-disapp-events} we present the total number of
CC $\nu_e$ scattering events on $^{12}$C in 5 to 50 kt LENA type detectors using the information on the source and
the detector characteristics from Table~\ref{tab:beamparams} and~\ref{tab:detector} respectively. Column 4 and 5
show the number of survived events after oscillation using the two different sets
of (3+2) parameter values given in Table~\ref{tab:global-3+2}. The last column shows the total event
rate without any oscillation. Here we have used total flux of $4 \times 10^{21}$ $\nu_e$ with a neutrino energy
threshold of 33 MeV and an efficiency of 80\%. Table~\ref{tab:nova-disapp-events} shows the same for
the 14 kt NO$\nu$A far detector but with a neutrino energy threshold of 38 MeV and an efficiency of 50\%.
We can expect $\sim$ thirty-four thousand events in NO$\nu$A with no oscillation considering only
the contribution of the transition of $^{12}$C to the $^{12}$N ground state~\cite{Auerbach:2001hz}.
From Table~\ref{tab:ls-disapp-events} we can see that the impact of disappearance on the total event rate is
6.3\%/23.4\% for the A/B parameter sets in the 50 kt LENA detector.
One should note that this is not a mere counting experiment
and that the $L/E$ pattern of the oscillations provides most of the measurement
sensitivity. This can be seen in Fig.~\ref{fig:ls-disapp-spec} where we plot the ratio of events
with and without sterile oscillation as a function of the reconstructed $L/E$
in the 50 kt LENA detector. With an L range of 20 - 120 m and with an energy range of 33 - 52.8 MeV, 
the $L/E$-dependence of the oscillation pattern can be
well observed inside the detector. The solid red line is for the oscillation parameter values in the (3+2) model as
given in the upper row of Table~\ref{tab:global-3+2}. The blue dashed line
shows the results for the (3+1) best fit values: $\Delta m^2_{41}$ = 1.78 eV$^2$ and $\ts_{ee}$ = 0.089
obtained from reactor anti-neutrino data with the new predictions for the reactor flux~\cite{Kopp:2011qd}. 
As seen in Fig.\ref{fig:ls-disapp-spec}, the shapes are quite different for (3+1) and (3+2) waves. Also,
a comparison between the amplitudes of the wave in various $L/E$ bins cancels flux normalization
and background systematic uncertainties to a large extent.

%%%%%%%%%%%%%%%%%%%%%%%%%%%%%%%%%%%%%%%%%%%%%%%%%%%%%%%%
%%%%%%%%%%%%%%%%%%%%%%%%%%%%%%%%%%%%%%%%%%%%%%%%%%%%%%%%%

%%%%%%%%%%%%%%%%%%%%%%%%%%%%%%%%%%%%%%%%%%%%%%%%%%%
\begin{table}[tp]
\begin{center}
  \begin{tabular}{||c||c||c||c||} \hline \hline
    Fiducial Mass & Flux & Flux \\
                  &  (A : Ref. \cite{Kopp:2011qd})   &  (B : Ref. \cite{Karagiorgi:2009nb}) \\
    \hline
    50 kt & $3.39 \times 10^{20}$ & $0.093 \times 10^{20}$ \\
    \hline
    25 kt & $5.55 \times 10^{20}$ & $0.214 \times 10^{20}$ \\
    \hline
    10 kt & $11.25 \times 10^{20}$ & $0.569 \times 10^{20}$ \\
    \hline
    5 kt & $22.1 \times 10^{20}$ & $1.202 \times 10^{20}$ \\
    \hline \hline
\end{tabular}
\caption{\label{tab:ls-disapp-flux} It shows the amount of neutrino flux needed to exclude 
the two different sets of oscillation parameter values as given in Table~\ref{tab:global-3+2} 
at $3\,\sigma$ CL with 4 degrees of freedom ($\Delta\chi^2 = 16.25$) using 5 to 50 kt LENA 
in disappearance mode.} 
\end{center}
\end{table}
%%%%%%%%%%%%%%%%%%%%%%%%%%%%%%%%%%%%%%%%%%%%%%%%%%

%%%%%%%%%%%%%%%%%%%%%%%%%%%%%%%%%%%%%%%%%%%%%%%%%%%
\begin{table}[tp]
\begin{center}
  \begin{tabular}{||c||c||c||c||} \hline \hline
    Total & Flux & Flux \\
   Background &  (A : Ref. \cite{Kopp:2011qd})   &  (B : Ref. \cite{Karagiorgi:2009nb}) \\
    \hline
     50000 & $5.9 \times 10^{21}$ & $1.325 \times 10^{21}$ \\
    \hline
     25000 & $4.615 \times 10^{21}$ & $0.963 \times 10^{21}$ \\
    \hline
     10000 & $3.408 \times 10^{21}$ & $0.636 \times 10^{21}$ \\
    \hline
      0    & $1.742 \times 10^{21}$ & $0.0945 \times 10^{21}$ \\
    \hline \hline
\end{tabular}
\caption{\label{tab:nova-disapp-flux} It depicts the amount of neutrino flux needed to exclude
the two different sets of oscillation parameter values as given in Table~\ref{tab:global-3+2} at $3\,\sigma$
CL with 4 degrees of freedom ($\Delta\chi^2 = 16.25$) using 14 kt NO$\nu$A far detector in disappearance 
mode. We show the required amount of flux for four different choices of effective Michel electron 
background.}
\end{center}
\end{table}
%%%%%%%%%%%%%%%%%%%%%%%%%%%%%%%%%%%%%%%%%%%%%%%%%%

In Table~\ref{tab:ls-disapp-flux} we present the amount of $\nu_e$ flux that is needed in
5 to 50 kt LENA type detectors
to exclude the two different sets of oscillation parameter values in the (3+2) scheme
as given in Table~\ref{tab:global-3+2}. The results are presented at $3\,\sigma$ CL (4 dof)
under the assumption that
there are no beam or non-beam backgrounds for LENA disappearance
search (see \Sec~\ref{sec:lena}). In this analysis, we consider 15\% systematic error on the signal and we follow the same
numerical method used for the appearance results.
Table~\ref{tab:ls-disapp-flux} shows that the required amount of flux is highly dependent on the
choice of the parameter values rather than the detector size. For example, we need
$\sim$ 26 times more flux for the parameter set A compared to B in the
25 kt LENA option. In Table~\ref{tab:nova-disapp-flux}, we present the results for NO$\nu$A considering a
20\% systematic error on signal and a 5\% systematic error on Michel decay backgrounds as measured 
during beam-off running. From this table, one can see the impact of the different choices of effective background 
on the required flux as compared to the no-background case.

%%%%%%%%%%%%%%%%%%%%%%%%%%%%%%%%%%%%%%%%%%%%%%%%%%%%%%%%
\begin{figure}[]
\centering
\includegraphics[width=0.5\textwidth]{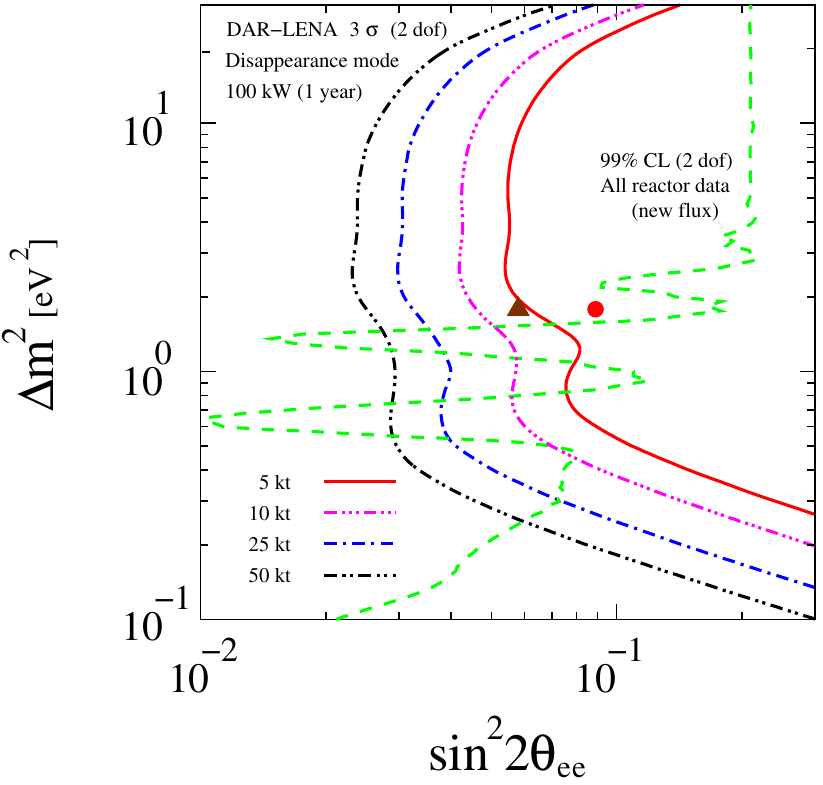}
\mycaption{\label{fig:ls-disapp-sens} Expected constraints from DAR-LENA setup on sterile neutrinos 
in the (3+1) model at $3\,\sigma$ CL (2 dof) using disappearance search. The dashed green curve shows the
99\% CL (2 dof) limit from reactor anti-neutrino data with new reactor fluxes~\cite{thomas}.  
The triangle and the bullet show the (3+1) best-fit values of all reactor data with old and new fluxes
respectively. Results are shown for 5 to 50 kt LENA type detector with $4 \times 10^{21}$ $\nu_e$.}
\end{figure}
%%%%%%%%%%%%%%%%%%%%%%%%%%%%%%%%%%%%%%%%%%%%%%%%%%%%%%%%%

%%%%%%%%%%%%%%%%%%%%%%%%%%%%%%%%%%%%%%%%%%%%%%%%%%%%%%%%
\begin{figure}[]
\centering
\includegraphics[width=0.5\textwidth]{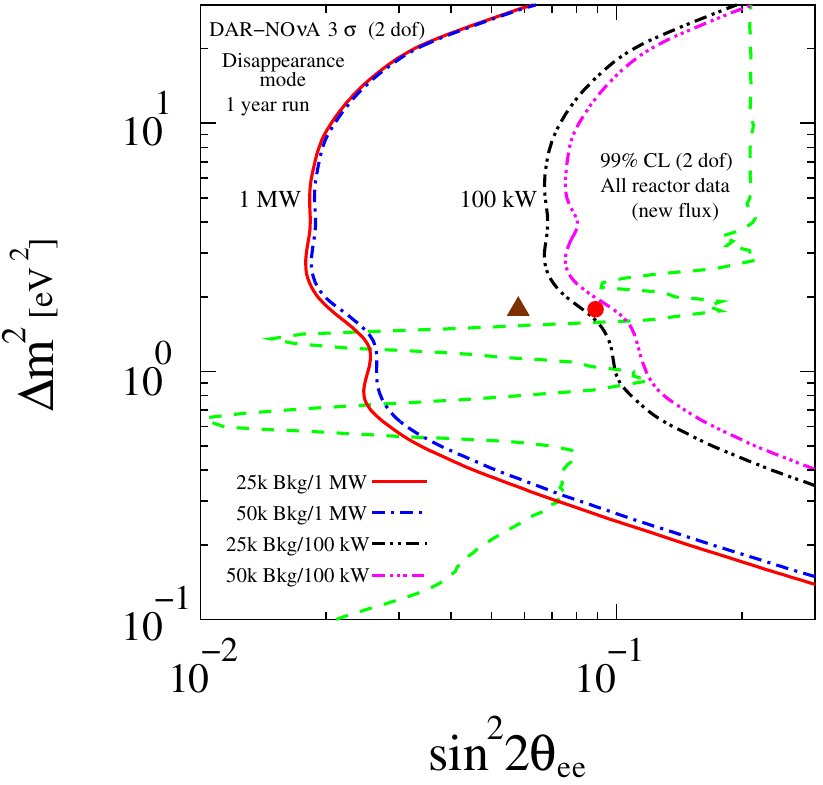}
\mycaption{\label{fig:nova-app-sens} Expected sensitivity limit of DAR-NO$\nu$A setup to (3+1) sterile neutrino 
oscillation at $3\,\sigma$ CL (2 dof) in disappearance mode. The dashed green line shows the
99\% CL (2 dof) limit from reactor anti-neutrino data with new reactor fluxes~\cite{thomas}.  
The triangle and the bullet show the (3+1) best-fit values of all reactor data with old and new fluxes
respectively. Results are presented considering 25000 and 50000 effective Michel electron backgrounds
with $4 \times 10^{21}$ (100 kW) and $4 \times 10^{22}$ (1 MW) $\nu_e$.}
\end{figure}
%%%%%%%%%%%%%%%%%%%%%%%%%%%%%%%%%%%%%%%%%%%%%%%%%%%%%%%%%

In Fig.~\ref{fig:ls-disapp-sens} we show the sensitivity limit of the
DAR-LENA setup to sterile neutrinos
in a (3+1) model at 3$\,\sigma$ CL (2 dof) using the disappearance channel. We compare our results
with the 99\% CL (2 dof) limit from reactor anti-neutrino data with the new reactor fluxes~\cite{thomas}.
The triangle ($\Delta m^2_{41}$ = 1.78 eV$^2$ and $\ts_{ee}$ = 0.058) and the
bullet ($\Delta m^2_{41}$ = 1.78 eV$^2$ and $\ts_{ee}$ = 0.089) show the (3+1) best-fit values of all reactor data
with the old and new fluxes respectively. These are shown as benchmark points to judge the performance of
the LENA  setup. A 10 kt LENA with a flux of $4 \times 10^{21}$ $\nu_e$ is sufficient to cover these test points and can
provide stringent test of the recent reactor anomaly with high significance.
Fig.~\ref{fig:nova-app-sens} shows
the sensitivity limit of the DAR-NO$\nu$A setup to sterile neutrinos in a (3+1) model.
Again, results are presented at 3$\,\sigma$ CL (2 dof) for the disappearance mode.
We show the results considering 25000 and 50000 effective Michel electron backgrounds
with $4 \times 10^{21}$ (100 kW) and $4 \times 10^{22}$ (1 MW) $\nu_e$ total fluxes.
For NO$\nu$A, a 100 kW machine is marginal in covering the test points and
a higher-power, full DAE$\delta$ALUS type machine, is needed to cover the entire parameter space.
By comparing Fig.~\ref{fig:ls-disapp-sens} and Fig.~\ref{fig:nova-app-sens},
one  can say that a 14 kt NO$\nu$A with 10 times larger flux can give similar performance
as to a 50 kt LENA with a reference flux of $4 \times 10^{21}$ $\nu_e$.

%%%%%%%%%%%%%%%%%%%%%%%%%%%%%
\section{Summary and Conclusions}
\label{sec:conclusion}
%%%%%%%%%%%%%%%%%%%%%%%%%%%%%

A host of recent SBL neutrino oscillation experiments have provided
hints that there may be oscillations to sterile neutrinos with a mass
squared difference of the order 0.1--10 eV$^2$. Sterile neutrino
models involving three active and one (3+1) or two (3+2) sterile
neutrino states have been proposed to explain these results. These
models demand that there be both appearance and disappearance
associated through or to sterile neutrinos. Thus, to address the
validity of these sterile neutrino models, better precision
measurements of both appearance and disappearance oscillations in this
mass region will be needed.

The combination of a long liquid-scintillator neutrino detector
combined with a cyclotron DAR neutrino source is very sensitive to
neutrino oscillations for both the $\bar \nu_e$ appearance and the
$\nu_e$ disappearance channels.  Such an experiment can observe the
$L/E$ variation of the oscillation rate within the detector and,
therefore, provide proof that the data is explained by neutrino
oscillations rather than other types of models. In addition, using the
$L/E$ variations within the experiment makes these types of
measurements fairly insensitive to normalization uncertainties and
backgrounds that do not have the $L/E$ dependence of neutrino
oscillations.

The future LENA experiment provides an example of the capabilities of
a very-large, liquid-scintillation detector. As shown above, the LENA
detector provides unmatched sensitivity in the $\bar \nu_e$ appearance
channel and completely covers the LSND and MiniBooNE signal regions at
more than 5~$\sigma$ confidence level with a cyclotron power of 10 kW
combined with a 5 kt detector. This would be a definitive
investigation of the the LSND/MiniBooNE reported signal and the
distinct $L/E$ dependence of the appearance signal would provide
unique confirmation that the signal is associated with high-$\Delta
m^2$ oscillations.

In addition, the LENA experiment could provide a robust search for
$\nu_e$ disappearance by observing the $L/E$ dependence of the
detected electron neutrinos. The sensitivity of the experiment would
cover the region suggested by the recent reactor $\bar \nu_e$
disappearance observation at 3~$\sigma$ and, again, would provide key
information on the possible interpretation of the result through the
$L/E$ dependence of the disappearance rate.

On an earlier timescale, the NO$\nu$A detector, now under
construction, combined with a 100 kW neutrino source could provide an
initial test of $\nu_e$ disappearance at high $\Delta m^2$ with mixing
angle sensitivity in the range of the recent reactor result.  Since
such a measurement is dominated by statistical uncertainties, making a
NO$\nu$A measurement with a 1 MW source would have a 3~$\sigma$
sensitivity at small mixing angle down to $\sin^22\theta_{ee} \sim
0.02$.

In conclusion, we have shown that large neutrino detectors using
liquid scintillator combined with high intensity 10-100 kW cyclotron
DAR neutrino sources would have unprecedented sensitivity to sterile
neutrino oscillations in the 0.5-10 eV$^2$ mass squared difference
region.  Such experiments are therefore an important option as a next
step in investigating neutrino oscillations to sterile neutrinos.

%%%%%%%%%%%%%%%%%% 
\subsubsection*{Acknowledgments}
%%%%%%%%%%%%%%%%%%

We thank P. Hernandez, P. Huber, M. Messier, L. Oberauer,
R.S. Raghavan, and T. Schwetz for useful discussions.
S.K.A. acknowledges the support from the European Union under the
European Commission Framework Programme~07 Design Study EUROnu,
Project 212372 and the project Consolider-Ingenio CUP.  J.M.C. and
M.H.S. are supported by the National Science Foundation.

\clearpage
\bibliographystyle{apsrev}
\bibliography{references}

\end{document}